\documentclass[aip,jcp,twocolumn,amsmath,amssymb]{revtex4}

\usepackage[english]{babel}
\usepackage{graphicx}
\usepackage{bm}

\begin{document}

\preprint{}

\title{Imaginary Time Correlations and the phaseless Auxiliary Field Quantum Monte Carlo}

\author{M. Motta, D.E. Galli}
\affiliation{$\mbox{Dipartimento di Fisica, Universit\`a degli Studi di Milano, via Celoria 16, 20133 Milano, Italy}$}
\author{S. Moroni}
\affiliation{IOM-CNR DEMOCRITOS National Simulation Center and SISSA, via Bonomea 265, 34136 Trieste, Italy}
\author{E. Vitali}
\affiliation{$\mbox{Dipartimento di Fisica, Universit\`a degli Studi di Milano, via Celoria 16, 20133 Milano, Italy}$ \\ $\phantom{3}$}
\date{\today}

\begin{abstract}

The phaseless Auxiliary Field Quantum Monte Carlo method provides a well established approximation scheme for accurate calculations of ground state energies of many-fermions systems.
Here we apply the method to the calculation of imaginary time correlation functions.
We give a detailed description of the technique and we test the quality of the results for static and dynamic properties against exact values for small systems.

\end{abstract}
 

\maketitle

\newcommand{\beq}{\begin{equation}}
\newcommand{\eeq}{\end{equation}}
\newcommand{\ket}[1]{| #1 \rangle}
\newcommand{\bra}[1]{\langle #1 |}
\newcommand{\braket}[2]{\langle #1 | #2 \rangle}
\newcommand{\tr}[1]{\mbox{tr}\left[ #1 \right]}
\newcommand{\mat}[2]{\mathcal{#1}_{#2}}
\newcommand{\dop}{\hat{\rho}}
\newcommand{\he}{${}^4 \mbox{He}$}
\newcommand{\pvm}[1]{\hat{\Pi}_{#1}}
\newcommand{\hs}{$\mathcal{H}\,$}
\newcommand{\choi}[2]{\ket{#1}\bra{#2}}
\newcommand{\crt}[1]{\hat{a}^\dag_{#1}}
\newcommand{\dst}[1]{\hat{a}_{#1}}
\newcommand{\ham}{\hat{H}}
\newcommand{\evo}[1]{e^{- #1 \ham}}
\newcommand{\gofeta}[1]{\hat{G}(\boldsymbol{\eta}_{#1})}
\newcommand{\gshift}[1]{\hat{G}(\boldsymbol{\eta}_{#1}-\boldsymbol{\xi}_{#1})}
\newcommand{\tns}[3]{\mathcal{#1}^{i_1 \dots i_{#2}}_{\phantom{i_1 \dots i_{#2} \, }j_1 \dots j_{#3}}}
\newcommand{\ster}{\theta,\phi}
\newcommand{\harm}[2]{Y_{#1,#2}(\ster)}
\newcommand{\qzero}{\ket{0}}
\newcommand{\qone}{\ket{1}}
\newcommand{\dt}{\delta\tau}
\newcommand{\psit}{\ket{\Psi_T}}
\newcommand{\eiva}[1]{\epsilon_{#1}}
\newcommand{\eive}[1]{\ket{\Phi_{#1}}}
\newcommand{\evodmc}[1]{e^{- #1 \left(\ham-\eiva{0}\right)}}
\newcommand{\slater}{$\mathfrak{D}(N)$}
\newcommand{\matel}[1]{\langle \hat{#1} \rangle}
\newcommand{\mateldue}[1]{\langle {#1} \rangle}

\section{INTRODUCTION}

\begin{figure}[b]
\label{fnvsexact}
\begin{center}
\includegraphics*[scale=0.35]{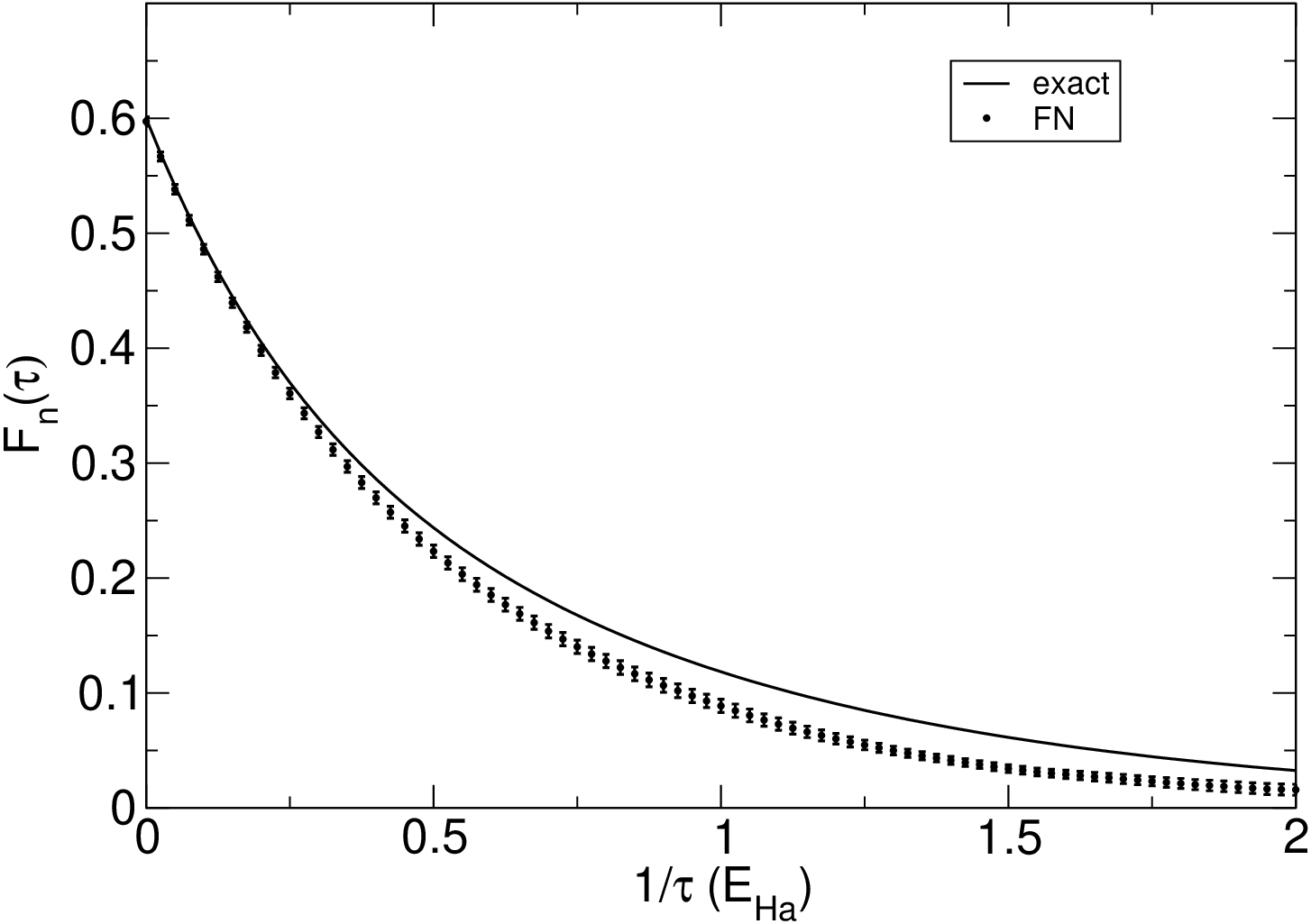}
\end{center}
\caption{FN result (points) and exact value (line) of the imaginary time correlation function 
of $\hat{\rho}_{\boldsymbol{q}}$ with $\boldsymbol{q} = \frac{2\pi}{L} (0,1)$ for a 2D system of $5$ noninteracting spinless fermions. 
Details of the calculation are presented in appendix \ref{appC}.}
\end{figure}

Over the last decades the study of many body quantum systems at zero temperature has been 
systematically supported by \emph{ab initio} Quantum Monte Carlo (QMC) calculations.
QMC are methods relying on a stochastic solution of the imaginary time Schr\"odinger 
equation of the system.
As far as \emph{bosonic} degrees of freedom are considered, QMC calculations allow static 
properties, energetics and structure functions, to be computed \emph{exactly}\cite{gfmc,cep,rept,pigs,spigs,patate} 
even for strongly correlated systems, for which analytic approaches yield only approximate results. Furthermore the 
possibility of reconstructing dynamical properties of bosonic 
systems, like excitation spectra and response functions, from imaginary time correlation functions 
has been explored with remarkable results\cite{gift5,rept,gift,gift8,gift9,gift10,gift11,overpress}.
On the other hand, for fermionic degrees of freedom the situation is considerably complicated by the well-known 
\emph{sign problem}\cite{feynman,loh}: computational cost increases exponentially with the system size.
The most widely employed scheme to reduce the problem to polynomial complexity is the \emph{Fixed-Node} (FN) 
approximation\cite{fn2,fn}: FN restricts the stochastic sampling of the \emph{configurational space} to regions where 
the sign of a reasonable approximation for the ground state wave function, the \emph{trial wave function}, remains constant.
Such approximation provides very accurate estimations of ground state properties\cite{fn2,fn,fci,fc}. 
Nevertheless, FN may give inaccurate results for imaginary time correlation functions even 
when the nodal structure of the ground state wavefunction is exactly known:
as an example, in Fig.~\ref{fnvsexact} we show the comparison between exact and 
FN imaginary time correlation functions of the density fluctuations $\langle \hat{\rho}_{\boldsymbol{q}}(\tau)
\hat{\rho}^{\dagger}_{\boldsymbol{q}} \rangle$ for a 2D system of $5$ noninteracting spinless fermions.
Such mismatch arises from the imposition of the ground state nodal structure as a subset of the nodal structure of all excited states. 
It is thus very interesting to investigate the possibility of extracting dynamical properties from QMC calculations within 
methods different from FN.

In recent years alternative QMC methods have been conceived, which 
simulate the imaginary time evolution with a suitable stochastic process taking place in 
the manifold of Slater determinants. \cite{af1,af3/2,af2,fci,fci_dyn}.
In the present work we consider one of such QMC methods, the 
phaseless Auxiliary Fields Quantum Monte Carlo (AFQMC) \cite{af1,af2,senechal},
which is considered less sensitive than FN to the quality of the trial function \cite{ceperley}.
However, the phaseless approximation is less known than those characterizing 
configuration QMC methods, and its accuracy in the calculation of imaginary 
time correlation functions is largerly unexplored.
In the present work we give a detailed description of AFQMC
and present its application to the calculation of imaginary time correlation functions.
To assess the accuracy of the phaseless AFQMC we compute static and
dynamic properties for a class of interacting fermionic models
amenable to exact diagonalization of the hamiltonian.
We will also compare AFQMC and FN results
for imaginary time correlation functions of larger systems.

The phaseless AFQMC and its extension to the calculation of dynamic properties are described in section I.
The solvable fermionic systems are presented in section II.
Results of numeric calculations are presented in section III, and conclusions are drawn in section IV.

\section{The Phaseless AFQMC}

As mentioned in the introduction, Quantum Monte Carlo are \emph{ab initio}: this means
that the starting point is the Hamiltonian operator of a physical system.
We thus present the phaseless AFQMC relying on a very general hamiltonian operator:

\begin{equation}
\label{hamiltonian}
\ham = \sum_{i,j=1}^M \beta_{ij} \, \hat{a}^\dag_i \hat{a}_j + \sum_{i,j,k,l=1}^M \gamma_{ijlk} \, \hat{a}^\dag_i \hat{a}^\dag_j \hat{a}_k \hat{a}_l
\end{equation}
The creation and destruction operators appearing in \eqref{hamiltonian}
are related to an orthonormal complete set of orbitals $\left\{ \ket{\varphi_i} \right\}_{i=1}^M$
in the single-particle Hilbert space, which we will denote \hs. $M$ is the dimension of
such Hilbert space; we will make the assumption that $M < +\infty$.
The above written Hamiltonian operator acts on the fermionic Fock space, $\mathcal{F}$,
built upon the one body space \hs.
Throughout this paper, we will fix the number of particles $N$, which is a constant
of motion for \eqref{hamiltonian}. Within the $N$--particles subspace of 
the Fock space $\mathcal{F}$, the operator $\hat{H}$ has the spectral resolution:

\begin{equation}
\label{spectral}
\ham = \sum_{\alpha} \eiva{\alpha} \, \ket{\Phi^{}_{\alpha}} \bra{\Phi_{\alpha}}
\end{equation}
where $\eiva{\alpha}$ are the eigenvalues and $\ket{\Phi^{}_{\alpha}}$ the eigenvetors.
Naturally the above expression is $N$--dependent, but we will not include
an explicit label $N$ to simplify the notation. The sum over $\alpha$ ranges from $0$ to the dimension of the $N$--body fermionic space, equal to $\binom{M}{N}$.
While zero temperature equilibrium properties of an $N$-particle system are completely determined by the 
ground state of \eqref{hamiltonian}, $\ket{\Phi^{}_{0}}$, the study of dynamic properties 
requires knowledge of the spectrum $\{\eiva{\alpha}\}_{\alpha}$. 
Throughout the present work we shall make the technical assumption that 
$\eiva{0} < \eiva{1} \leq \dots $ i.e. that the $N$-particle ground state is non-degenerate.

A wide class of QMC methods relies on the observation that the imaginary time propagator:
\begin{equation}
\label{utau}
\evo{\tau}, \quad \tau \geq 0
\end{equation}
enables the ground state of an $N$-particle system to be recovered. 
In fact, as long as a trial state $\psit$ has non--zero overlap with 
$\eive{0}$ the following relation holds:
\begin{equation}
\label{evolution}
\lim_{\tau \to \infty} \evodmc{\tau} \psit = \eive{0} \braket{\Phi_0}{\Psi_T}
\end{equation}
where the unknown quantity $\eiva{0}$ is replaced with an \emph{adaptive} estimate, according to a common procedure in DMC calculations \cite{fn}.
QMC methods rely on the observation that deterministic evolution driven by the family of operators \eqref{utau} can be mapped onto suitable stochastic processes and solved by randomly sampling appropriate probability distributions.

Along with the typical approach in which \eqref{evolution} is associated to a diffusion process in the configurational space of the system\cite{pigs,spigs,fn}, in a class of more recently developed QMC methods, the so-called \emph{determinantal}\cite{fc,fci,af1,af3/2,af2} methods, \eqref{evolution} is mapped onto a stochastic process in the abstract manifold, which we will
denote \slater, of $N$-particle Slater determinants. 

In AFQMC, first conceived by G. Sugiyama and S. E. Koonin \cite{af1}, later perfected 
and extended by S. Zhang\cite{af3/2,af2,senechal} and F.Assaad\cite{assaad} and successfully applied to the investigation of molecular systems \cite{jcpshiwei1,jcpshiwei2,jcpshiwei3}, the association 
between \eqref{evolution} and a stochastic process in \slater~ is made possible by a discretization:
\begin{equation}
\label{dt_def}
\evodmc{\tau} =  \left( \evodmc{\dt} \right)^n
\end{equation}
with $\dt = \frac{\tau}{n}$, and by a combined use of the Trotter-Suzuki decomposition of the propagator\cite{trotter,suzuki} and of the \emph{Hubbard-Stratonovich transformation} \cite{hubbard,stratonovich,senechal} on the factors $\evodmc{\dt}$. The Hubbard-Stratonovich transformation is an operator identity guaranteeing that:
\begin{equation}
\label{hs-transform}
\evodmc{\dt} = \int dg(\boldsymbol{\eta}) \, \hat{G}(\boldsymbol{\eta}) + \mathcal{O}(\dt^2)
\end{equation}
with $dg(\boldsymbol{\eta})$ standard $2 M^2$-dimensional normal probability measure, $\hat{G}(\boldsymbol{\eta}) = e^{\hat{A}(\boldsymbol{\eta})}$ and $\hat{A}(\boldsymbol{\eta}) = \sum_{i,j=1}^M \mathcal{A}(\boldsymbol{\eta})_{i,j} \crt{i} \dst{j}$ a suitable \emph{one-particle operator}, the structure of which is discussed in detail in \ref{appA2}.

Equation \eqref{hs-transform} establishes a formal correspondence between an interacting fermion system and an ensemble of \emph{non-interacting} fermion systems subject to \emph{fluctuating external potentials}. The coupling with these external potentials is controlled by normally-distributed parameters $\boldsymbol{\eta}$, called \emph{ auxiliary fields}, integration over which recovers the interaction.

To quantitatively realize that \eqref{hs-transform} provides a random walk representation of the imaginary time evolution, let us consider the stochastic process defined by the succession of wave functions:
\begin{equation}
\label{stochastic}
\ket{\Psi_n} = \hat{G}(\boldsymbol{\eta}_{n-1}) \dots \hat{G}(\boldsymbol{\eta}_{0}) \psit
\end{equation}
where the operators $\hat{G}(\boldsymbol{\eta}_k)$ are functions of independent normally-distributed random variables $\boldsymbol{\eta}_k$. 
It is known, and will be shown in details in \ref{appA1}, that, if $\psit \in \mathfrak{D}(N)$, 
all the random variables $\ket{\Psi_n}$ take values in $\mathfrak{D}(N)$. 
Furthermore, their average is given by:
\begin{equation}
\label{stochastic2}
\begin{split}
&\langle \ket{\Psi_n} \rangle_{\boldsymbol{\eta}_{n-1} \dots \boldsymbol{\eta}_0} = \langle \hat{G}(\boldsymbol{\eta}_{n-1}) \dots \hat{G}(\boldsymbol{\eta}_{0})  \rangle_{\boldsymbol{\eta}_{n-1} \dots \boldsymbol{\eta}_0} \psit = \\
= &\int dg(\boldsymbol{\eta}_{n-1}) \dots dg(\boldsymbol{\eta}_0) \, \hat{G}(\boldsymbol{\eta}_{n-1}) \dots \hat{G}(\boldsymbol{\eta}_0) \psit = \\
= \, &\evodmc{n\dt} \psit + \mathcal{O}(\dt^2) \\
\end{split}
\end{equation}
The expression \eqref{stochastic2} clearly shows that the solution of the imaginary time 
Schr\"odinger equation \eqref{evolution} can be recovered as average of a suitable stochastic 
process, the structure of which is suggested by \eqref{stochastic}.
Combining \eqref{stochastic} and \eqref{evolution} it is evident that 
numerical sampling of such stochastic process provides a stochastic linear combination of Slater determinants, representing an estimation of the ground state of \eqref{hamiltonian}.

\subsection{control of the fermion sign problem: \\the phaseless AFQMC}

Although its formal simplicity, the straightforward numerical implementation
leads in general to an exponential increase in statistical errors with
the imaginary time, due to the fact that complex random phases appear during
the evolution \eqref{evolution}.

S. Zhang invented a stabilization procedure to modify the stochastic process
in order to plug into the sampling information that \emph{guides}
the random walk, closely resembling the typical scheme adopted
in configurational DMC simulations: 
an \emph{importance sampling transformation}
\cite{senechal}. 
The state $\evodmc{n \dt} \psit$ is rewritten in the following form, detailed in \ref{appA3} and equivalent to \eqref{stochastic2}:
\begin{eqnarray}
\label{randomwalk}
\nonumber
&\evodmc{n \dt} \psit \simeq \int dg(\boldsymbol{\eta}_{n-1}) \dots dg(\boldsymbol{\eta}_0) \, \\ 
\nonumber
& \mathfrak{W}\left[ \boldsymbol{\eta}_{n-1}, \boldsymbol{\xi}_{n-1} \dots \boldsymbol{\eta}_0, \boldsymbol{\xi}_0 \right]  \, \frac{\gshift{n-1} \dots \gshift{0} \psit}{\braket{ \Psi_T}{\gshift{n-1} \dots \gshift{0} | \Psi_T}}\\
\nonumber
\\
\end{eqnarray}
where \emph{complex-valued shift parameters} $\boldsymbol{\xi}_{n-1} \dots \boldsymbol{\xi}_0$ and a weight function have been inserted. The latter satisfy the recursion relation:
\begin{eqnarray}
\label{weight}
\nonumber
&\mathfrak{W}\left[ \boldsymbol{\eta}_n, \boldsymbol{\xi}_n \dots \boldsymbol{\eta}_0, \boldsymbol{\xi}_0 \right] = \mathfrak{W}\left[ \boldsymbol{\eta}_{n-1}, \boldsymbol{\xi}_{n-1} \dots \boldsymbol{\eta}_0, \boldsymbol{\xi}_0 \right] \times \\
\times &\mathfrak{I}\left[ \boldsymbol{\eta}_n, \boldsymbol{\xi}_n ; \gshift{n-1} \dots \gshift{0} \psit \right]
\end{eqnarray}
where the following \emph{importance function}:
\begin{equation}
\label{importance}
\mathfrak{I}\left[ \boldsymbol{\eta}, \boldsymbol{\xi} ; | \Psi \rangle \right] = e^{-\frac{\boldsymbol{\xi} \cdot \boldsymbol{\xi}}{2} - \boldsymbol{\eta} \cdot \boldsymbol{\xi}} \frac{\langle \Psi_T | \hat{G}(\boldsymbol{\eta}-\boldsymbol{\xi}) | \Psi \rangle}{\langle \Psi_T | \Psi \rangle}
\end{equation}
appears. The shift parameters are chosen to minimize fluctuations in the importance function to first order in $\dt$. As it will be described in \ref{appA3}, the complex-valued importance function is subsequently turned into a \emph{real positive quantity} by the so-called \emph{real local energy}\cite{af2} approximation:
\begin{equation}
\label{rle}
\mathfrak{I}\left[ \boldsymbol{\eta}, \boldsymbol{\xi} ; | \Psi \rangle \right] \simeq e^{-\dt (\epsilon_{loc}(\Psi)-\epsilon_0)}
\end{equation}
where $\epsilon_{loc}(\Psi) = \mbox{re}\left[ \frac{\braket{\Psi_T}{\ham|\Psi}}{\braket{\Psi_T}{\Psi}} \right]$ is the \emph{local energy} functional. The importance sampling expressions \eqref{importance}, \eqref{rle} clearly show the mechanism responsible for the appearence of the sign problem in the framework of AFQMC: \emph{when the overlap between one or more walkers and the trial state vanishes} massive fluctuations in the importance function occur, determining \emph{drastic statistic errors in AFQMC estimates} \eqref{estimate}, \eqref{gs_average} for the solution of \eqref{evolution} and for ground state averages of many body observables.

Since it has been argued \cite{af2,senechal} that the vanishing of such overlap always occurs when its phase changes abruptly, control of the fermion sign problem is accomplished implementing the so-called \emph{phase approximation}, in which the importance function of walkers undergoing an abrupt phase change, in the sense that the quantity:
\begin{equation}
\Delta \theta = \mbox{Im} \left[ \log\left[ \frac{\braket{\Psi_T}{\Psi^{(w)}_{n+1}}}{\braket{\Psi_T}{\Psi^{(w)}_n}} \right] \right]
\end{equation}
has negative cosine, is put equal to 0.

We observe here that to our knowledge there is no rigorous proof that a perfect correlation relates sudden phase change and vanishing of the overlap with the trial state, and that the real local energy approximation and the phase approximation produce unbiased estimators \eqref{estimate}, \eqref{gs_average}. One of the topics investigated in the present work is the actual verification of these conditions for a model system.

\subsection{The Algorithm}

The so-far introduced observations give rise to a \emph{polynomially complex} algorithm for numerically sampling the solution \eqref{evolution}, the efficency of which relies on the observation that the walkers $\ket{\Psi^{(w)}_{k}}$ lie in $\mathfrak{D}(N)$ and can be therefore parametrized with an $M \times N$ complex-valued matrix, as discussed in \ref{appA1}. The algorithm can be resumed in the following sequence of operations:
\begin{enumerate}
\item a collection $\ket{\Psi^{(1)}_0} \dots \ket{\Psi^{(N_w)}_0}$ of Slater determinants, henceforth referred to as \emph{walkers}, is initialized to a trial state $\psit$.
\item for $k = 0 \dots n-1$ an adaptive estimate of the ground state energy is produced according to the formula:
\begin{equation}
\label{etrial}
\epsilon_0 \simeq \frac{1}{\sum_{w=1}^{N_w} \mathfrak{W}^{(w)}_{k}} \, \sum_{w=1}^{N_w} \mathfrak{W}^{(w)}_{k} \, \frac{\braket{\Psi_T}{\hat{H} |\Psi^{(w)}_{k}}}{\braket{\Psi_T}{\Psi^{(w)}_{k}}}
\end{equation}
normally distributed auxiliary field configurations $\boldsymbol{\eta}^{(1)}_k \dots \boldsymbol{\eta}^{(N_w)}_k$ are sampled, and walkers and weights are updated according to:
\begin{equation}
\begin{split}
\label{flow}
\ket{\Psi^{(w)}_{k+1}} &= \hat{G}(\boldsymbol{\eta}^{(w)}_k - \boldsymbol{\xi}^{(w)}_k) \, \ket{\Psi^{(w)}_{k}} \\
\mathfrak{W}^{(w)}_{k+1} &= \mathfrak{W}^{(w)}_{k} \, \mathfrak{I}\left[ \boldsymbol{\eta}^{(w)}_k , \boldsymbol{\xi}^{(w)}_k ; \ket{\Psi^{(w)}_{k}} \right]\\
\end{split}
\end{equation}
\item an estimate for $\evodmc{n \dt} \psit$ is given by:
\begin{equation}
\label{estimate}
\evodmc{n \dt} \psit \simeq \sum_{w=1}^{N_w} \mathfrak{W}^{(w)}_{n} \frac{\ket{\Psi^{(w)}_{n}}}{\braket{\Psi_T}{\Psi^{(w)}_{n}}}
\end{equation}
\end{enumerate}
The ground state average $\braket{\Phi_0}{\hat{O}|\Phi_0}$ of a many-body observable $\hat{O}$ not commuting with $\ham$ is the $m,n \to \infty$ limit of the following formula:
\begin{equation}
\label{backpro}
\frac{\braket{\Psi_T}{\evodmc{m \dt} \hat{O} \evodmc{n \dt}|\Psi_T}}{\braket{\Psi_T}{\evodmc{(m+n)\dt}|\Psi_T}}
\end{equation}
for which manipulations analogous to the importance sampling transformation, discussed in detail in \ref{appA4}, yield the following \emph{backpropagated} \cite{senechal} estimate:
\begin{equation}
\label{gs_average}
\braket{\Phi^{(N)}_0}{\hat{O}|\Phi^{(N)}_0} \simeq \frac{\sum_{w=1}^{N_w} \mathfrak{W}^{(w)}_{n+m} \frac{\langle \Psi^{(w)}_{BP,m}|\hat{O}|\Psi^{(w)}_{n} \rangle}{ \langle \Psi^{(w)}_{BP,m}|\Psi^{(w)}_n \rangle }}{\sum_{w=1}^{N_w} \mathfrak{W}^{(w)}_{n+m}}
\end{equation}
with:
\begin{equation}
\begin{split}
| \Psi^{(w)}_n \rangle &= \hat{G}(\boldsymbol{\eta}_{n-1}-\boldsymbol{\xi}_{n-1}) \dots \hat{G}(\boldsymbol{\eta}_0-\boldsymbol{\xi}_0) | \Psi_T \rangle \\
| \Psi^{(w)}_{BP,m} \rangle &= \hat{G}^\dag(\boldsymbol{\eta}_n-\boldsymbol{\xi}_n) \dots \hat{G}^\dag(\boldsymbol{\eta}_{n+m-1}-\boldsymbol{\xi}_{n+m-1}) | \Psi_T \rangle \\
\end{split}
\end{equation}

\subsection{Imaginary time correlation functions}

In a well-established approach\cite{gift5,gift8,gift9,gift10,gift11,fc,gift} to the reconstruction of dynamic properties of many body systems, the \emph{dynamic structure factor} of the single-particle operators $\hat{A},\hat{B}$:

\begin{equation}
\label{structure}
S_{\hat{A},\hat{B}}(\omega) = \int_{\mathbb{R}} dt \, \frac{e^{i\omega t}}{2 \pi} \, \braket{\Phi_0}{ \hat{A}(t)\hat{B} | \Phi_0}
\end{equation}

$ $

is recovered from their \emph{imaginary time correlation function} (ITCF):
\begin{equation}
\label{ITCF}
\begin{split}
F_{\hat{A},\hat{B}}(\tau) &= \frac{\braket{\Phi_0}{ \hat{A}(\tau)\hat{B} | \Phi_0}}{N} = \frac{\braket{\Phi_0}{ \hat{A} \evodmc{\tau} \hat{B} | \Phi_0}}{N} \\
\end{split}
\end{equation}
though a numeric inverse Laplace transform. Being constructed with the imaginary time evolution operator, the ITCF \eqref{ITCF} is a 
natural quantity to be evaluated in QMC calculations.
Its evaluation in determinantal QMC methods, however, is not as simple as in configurational QMC:
straightforward extension of the backpropagation technique to the evaluation of \eqref{ITCF} is in fact prevented because the single-particle operator $\hat{B}$ does not preserve \slater. To overcome this difficulty, we generalize the clever approach conceived by M. Feldbacher and F.Assaad\cite{assaad} for the calculation of dynamical Green function: we introduce the Hubbard-Stratonovich representation \eqref{hs-transform} of the imaginary time propagator in \eqref{ITCF} and move the operators $\hat{G}(\boldsymbol{\eta})$ to the right of $\hat{B}=\sum_{ij=1}^M \mathcal{B}_{ij} \, \crt{i}\dst{j}$ commuting them with the operators $\crt{i}$, $\dst{j}$. As discussed in detail in \ref{appA5}, this procedure determines the appearence of two random matrices in the estimator for \eqref{ITCF}. Concretely:
\begin{widetext}
\begin{equation}
\begin{split}
\label{ext_phaseless}
&F_{\hat{A},\hat{B}}(r \delta\tau) = \frac{1}{N} \sum_{kl} \, \mathcal{B}_{kl} \int dg(\boldsymbol{\eta}_{n-1}) \dots dg(\boldsymbol{\eta}_{n-r}) \braket{\Phi_0}{\hat{A} \gofeta{n-1} \dots \gofeta{n-r} \crt{k} \dst{l} |\Phi_0} = \\
= &\frac{1}{N} \sum_{ijkl} \mathcal{B}_{kl} \int dg(\boldsymbol{\eta}_{n-1}) \dots dg(\boldsymbol{\eta}_{n-r}) \braket{\Phi_0}{\hat{A} \crt{i} \dst{j} \gofeta{n-1} \dots \gofeta{n-r} |\Phi_0} \, \mathcal{D}(\boldsymbol{\eta}_{n-1} , \dots , \boldsymbol{\eta}_{n-r})_{ik} \, \mathcal{D}^{-1}(\boldsymbol{\eta}_{n-1} , \dots , \boldsymbol{\eta}_{n-r})_{lj} \\
\end{split}
\end{equation}
\end{widetext}
where $\mathcal{D}(\boldsymbol{\eta}_{n-1} , \dots , \boldsymbol{\eta}_{n-r}) = e^{\mathcal{A}(\boldsymbol{\eta}_{n-1})} \dots e^{\mathcal{A}(\boldsymbol{\eta}_{n-r})}$.

Further application of the importance sampling transformation and of the backpropagation technique yields, as explained in \ref{appA5}:

\begin{widetext}
\begin{equation}
\label{the_boss}
\begin{split}
F_{\hat{A},\hat{B}}(r \delta\tau) \simeq \frac{1}{N} \frac{1}{\sum_{w=1}^{N_w} \mathfrak{W}^{(w)}_{m+n-r}} \sum_{w=1}^{N_w} \sum_{ijkl} \, &\mathcal{B}_{kl} \, \mathfrak{W}^{(w)}_{m+n} \, \frac{\langle \Psi^{(w)}_{BP,m}|\hat{A} \hat{a}^\dag_i \hat{a}_j|\Psi^{(w)}_{n} \rangle}{\langle \Psi^{(w)}_{BP,m}|\Psi^{(w)}_n \rangle} \\
&\mathcal{D}(\boldsymbol{\eta}^{(w)}_{n-1} - \boldsymbol{\xi}^{(w)}_{n-1}, \dots , \boldsymbol{\eta}^{(w)}_{n-r} - \boldsymbol{\xi}^{(w)}_{n-r} )_{ik} \mathcal{D}^{-1}( \boldsymbol{\eta}^{(w)}_{n-1} - \boldsymbol{\xi}^{(w)}_{n-1}, \dots , \boldsymbol{\eta}^{(w)}_{n-r} - \boldsymbol{\xi}^{(w)}_{n-r} )_{lj} \\
\end{split}
\end{equation}
\end{widetext}

\section{a class of solvable systems}

We test the accuracy of the AFQMC results on a class of simple systems for which exact numeric expression for the spectral decomposition \eqref{spectral} of the Hamiltonian operator $\ham$ can be given.
Let us consider the Hamiltonian of the 2D electron gas,
\begin{equation}
\label{toy-hamiltonian}
\begin{split}
\ham &= \frac{\xi \sqrt{N}}{\sqrt{4\pi} r_s} + \sum_{m,\sigma} \frac{2 \pi}{N} \, \frac{|\boldsymbol{n}_m|^2}{r_s^2} \crt{m,\sigma} \dst{m,\sigma} + \\
&+\sum_{\sigma,\sigma'} \sum_{mnrs} \frac{1}{\sqrt{4 N \pi}\, r_s} \, \frac{\delta_{\boldsymbol{n}_r-\boldsymbol{n}_m,\boldsymbol{n}_n-\boldsymbol{n}_s}}{|\boldsymbol{n}_r-\boldsymbol{n}_m|} \crt{m,\sigma} \crt{n,\sigma'} \dst{s,\sigma'} \dst{r,\sigma} \\
\end{split}
\end{equation}
where the single-particle Hilbert space \hs is spanned by the plane wave orbitals $|\boldsymbol{n}_{i} \sigma \rangle$ with $\boldsymbol{n}_i \in \mathbb{Z}^2$, $|\boldsymbol{n}_i|^2 \leq n_{max}$ for some integer $n_{max}$ and $\sigma = \pm 1$. The parameter $r_s \in (0,\infty)$ controls the relevance of the interaction part and $N$ stands for the number of particles, and the constant $\xi = -3.900265$ arises from an Ewald summation procedure\cite{ewald}. For small number of particles $N$ and low kinetic energy cutoff $n_{max}$ the above Hamiltonian defines a simple model which can be solved exactly.

Knowledge of eigenvalues $\{ \eiva{\alpha} \}$ and eigenvectors $\{\eive{\alpha}\}$ of $\ham$ allows exact calculation of the imaginary time propagator:
\begin{equation}
\evodmc{\tau} = \sum_\alpha e^{-\tau \left( \eiva{\alpha} - \eiva{0} \right)} \eive{\alpha} \bra{\Phi_\alpha} \quad , \\
\end{equation}
of the projector $\eive{0} \bra{\Phi_0}$ onto the minimum energy eigenspace, of backpropagated ground state averages:
\begin{equation}
\begin{split}
&\frac{\braket{\Psi_T}{\evodmc{m\dt}\hat{O}|\Phi_0}}{ \braket{\Psi_T}{\evodmc{m\dt}|\Phi_0}} = \\
= &\frac{\sum_\alpha \braket{\Psi_T}{\Phi_\alpha} e^{-m\dt(\eiva{\alpha}-\eiva{0})} \braket{\Phi_\alpha}{\hat{O}|\Phi_0}}{\braket{\Psi_T}{\Phi_0}} \\
\end{split}
\end{equation}
and of backpropagated imaginary time correlation functions \eqref{ITCF}: 
\begin{widetext}
\begin{equation}
\label{exact_ITCF}
\begin{split}
F_{\hat{A},\hat{B}}(\tau)= \frac{1}{N} \frac{\braket{\Psi_T}{\evodmc{m\dt}\hat{A}\evodmc{r\dt}\hat{B}|\Phi_0}}{ \braket{\Psi_T}{\evodmc{m\dt}|\Phi_0}} = \frac{1}{N} \frac{\sum_{\alpha,\beta}\braket{\Psi_T}{\Phi_\alpha} e^{-m\dt(\eiva{\alpha}-\eiva{0})-r\dt(\eiva{\beta}-\eiva{0})} \braket{\Phi_\alpha}{\hat{A}|\Phi_\beta} \braket{\Phi_\beta}{\hat{B}|\Phi_0}  }{\braket{\Psi_T}{\Phi_0}} \\
\end{split}
\end{equation}
\end{widetext}
and the comparison of such quantities with AFQMC results. Particular importance shall be annected to the ITCF $F_{\boldsymbol{n}}(\tau) = \frac{1}{N} \braket{\Psi_0|\hat{\rho}_{\boldsymbol{n}}(\tau) \hat{\rho}_{-\boldsymbol{n}}}{\Psi_0}$ of the density fluctuation operator:
\begin{equation}
\label{mom_shift}
\hat{\rho}_{\boldsymbol{n}} = \sum_{i,j}\sum_{\sigma} \delta_{\boldsymbol{n}_i,\boldsymbol{n}_j-\boldsymbol{n}} \, \crt{i,\sigma} \dst{j,\sigma}
\end{equation}
and of its adjoint $\hat{\rho}_{\boldsymbol{n}}^\dag = \hat{\rho}_{-\boldsymbol{n}}$.

\begin{figure*}[ht!]
\label{fig_equil}
\begin{center}
\includegraphics[scale=0.6]{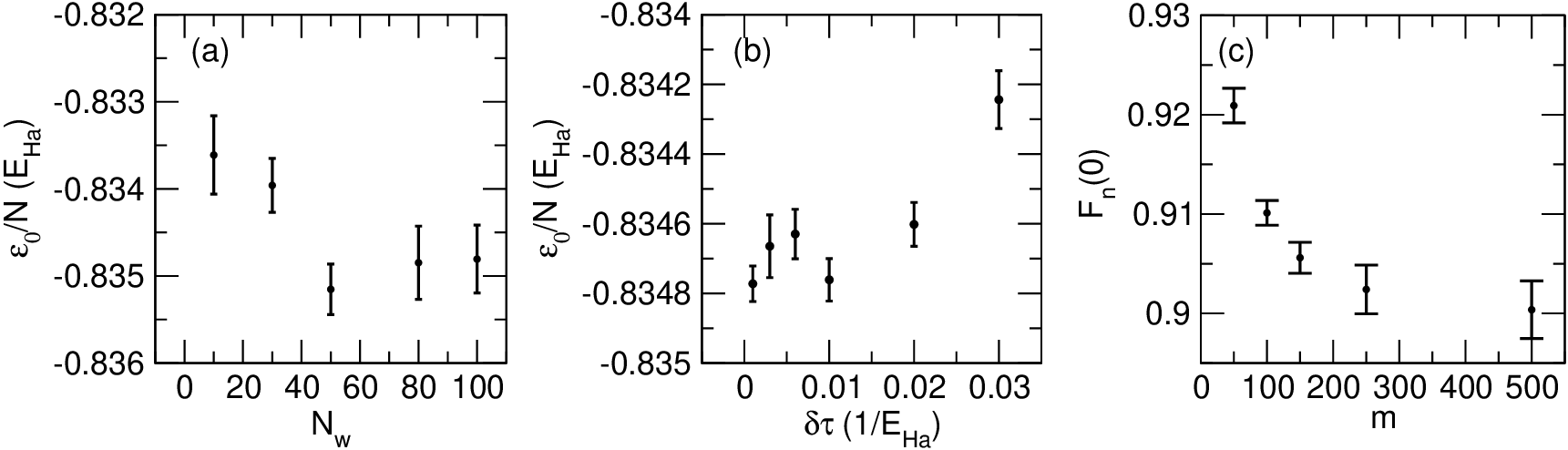}
\end{center}
\caption{Steps of the convergence procedure for $(N_\uparrow,N_\downarrow,r_s,M)= (1,1,1,49)$:
(a) AFQMC estimate of the ground state energy per particle for several values of $N_w$ at fixed 
$\delta\tau=0.001$: $N_w$ can be safely set to $80$ 
(b) AFQMC estimate of the ground state energy per particle for several values of $\delta \tau$ 
at fixed $N_w = 80$: $\delta \tau$ can be safely set to $0.003$ 
(c) AFQMC estimate of $F_{\bold{n}}(0)$ for several values of $m$ at fixed $N_w = 80$, 
$\delta\tau = 0.003$: $m$ can be safely set to $m=250$}
\end{figure*}

\section{Results} 

The phaseless AFQMC method represents the ground state as a stochastic linear combination of Slater determinants, \eqref{estimate}, from which accurate estimates of the ground state energy can be obtained \cite {af3/2,senechal}. However much more information can be obtained from the simulation. Here we present results for the components of the ground state on the chosen basis of the Hilbert space and for the imaginary time correlation functions.

Each of the simulations presented below is characterized by two sets of parameters: $(N_\uparrow,N_\downarrow,r_s,M)$ define the system under study, whereas $(\delta \tau, m,N_w)$ control the details of the simulation.
In particular, $N_\uparrow$ ($N_\downarrow$) is the number of spin-up (spin-down) fermions, $r_s$ controls the strength of the interaction, $M$ fixes the order of the matrices with which the algorithm deals, while $m$ corresponds to the number of backpropagation steps.

Apart from the basis set size $M$, which we keep small to allow comparison with exact diagonalization, we extrapolate to the joint limit $\delta\tau \to 0$, $m \to \infty$ and $N_w \to \infty$. As an example, we show in figure Fig.~\ref{fig_equil} the extrapolations for a calculation with $(N_\uparrow,N_\downarrow,r_s,M) = (1,1,1,49)$.
Discrepancies with respect to the exact results are therefore due to the uncontrolled approximations of the method, namely the \emph{real local energy} and the \emph{phase} approximations.

\subsection{Assessment of the accuracy}

In figure Fig.~\ref{(1,1,1,13)} we show results relative to the simulation of systems with $r_s=1$, $M = 13$, for some values of $N_\uparrow$ and $N_\downarrow$. 
The left panels of the figure show the components of the stochastic solution on the Hilbert space basis functions. 
The little statistical fluctuations around the $x$ axis show that the random walk visits a large number of states, 
while the significant components of the AFQMC solution match those of the exact ground state with good accuracy.
The ITCF $F_{\boldsymbol{n}}(\tau)$ of the density fluctuation operator \eqref{mom_shift} 
for $\bold{n}=(1,0)$ is reported in the right column of Fig.~\ref{(1,1,1,13)}.
The wave vector $\bold{n}$ has been chosen in the lowest energy shell since it gives rise to non-vanishing ITCFs even for small $M$.
The agreement with exact values is remarkable, in particular if compared with the discrepancy observed for the FN result, Fig.~\ref{fnvsexact}:
this constitutes the central result of the present work.

For all these systems we computed also the ground state energy per particle, 
and the overlap between exact and reconstructed ground state: the results are listed in table Tab.~\ref{ovr}, the bias of the energy resulting of the order of $10^{-3} E_{Ha}$, which is
smaller than the FN bias using a Slater-Jastrow trial function with plane-wave nodes \cite{kwon}.

\begin{figure*}
\label{(1,1,1,13)}
\centering
\includegraphics[scale = 0.6]{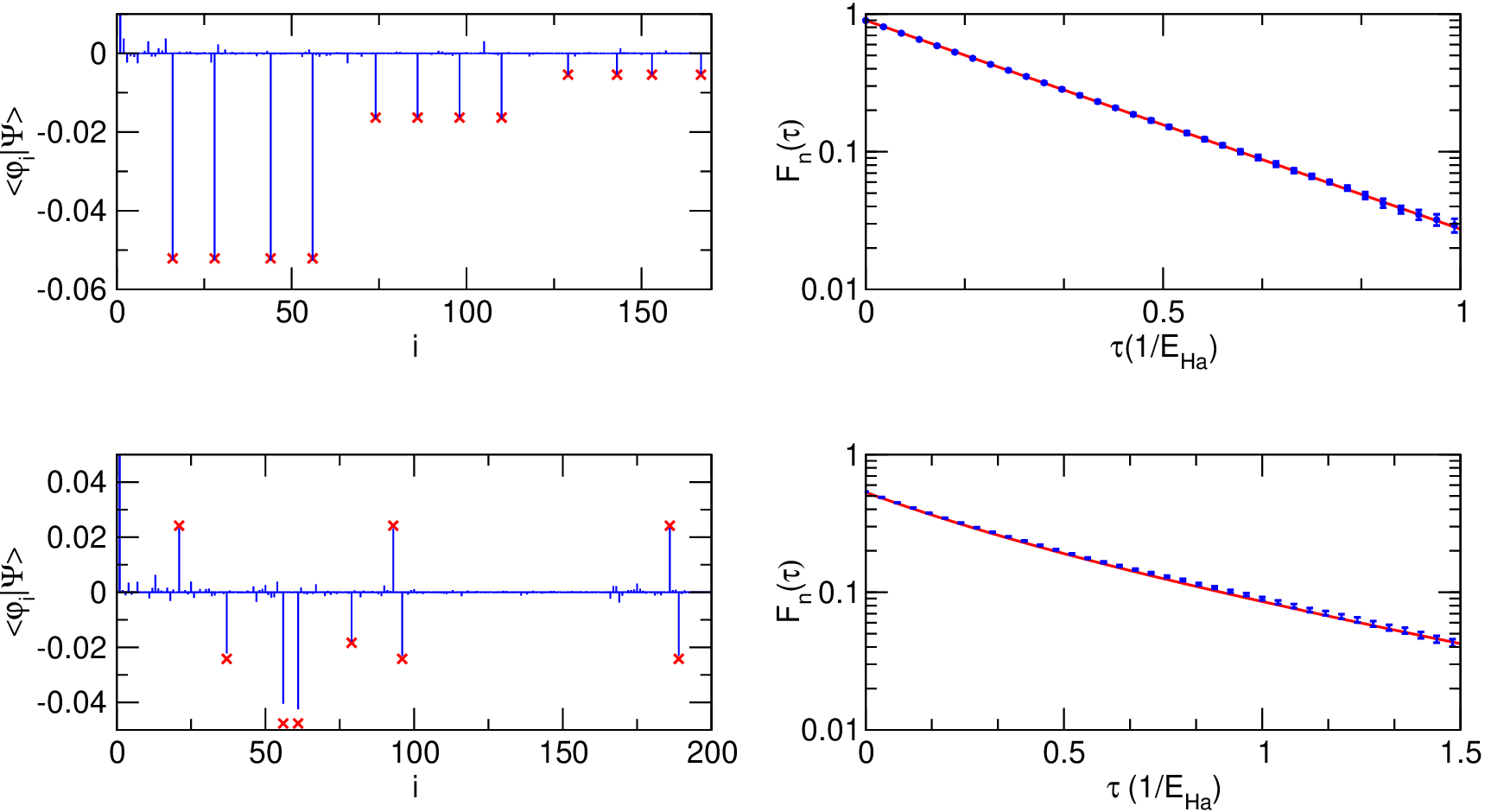}
\caption{(color online) Left column: exact nonvanishing (red crosses) and reconstructed (blue columns) components of the ground state for systems
with $N_\uparrow,N_\downarrow = (1,1),(5,0)$ (top to bottom), relative to all elements of the basis except $\ket{\varphi_1} = \ket{\Psi_T}$.
Exact and reconstructed values of $\braket{\varphi_1}{\Psi}$ are respectively $0.9937,0.9939(2)$ and $0.9926,0.9885(3)$.
Right column: exact (red line) and reconstructed (circles) imaginary time correlation function of the density fluctuation operator with
$\bold{n}=(1,0)$ for systems with $N_\uparrow,N_\downarrow = (1,1),(5,0)$ (top to bottom). When not visible, error bars are smaller than the symbol size.}
\end{figure*}

\pagebreak

\subsection{Computational Issues}

Although our primary interest is the assessment of the accuracy of AFQMC in calculating the ITCFs addressed in the previous section, 
we explored the behavior of the method for larger values of $r_s$ and $M$.

As $r_s$ increases, the interaction becomes more and more important, and the overlap of the exact wave function with the trial
function becomes smaller. Also the increase in $M$, which is required for the study of bulk systems, makes the stochastic exploration
of the Hilbert space more difficult: in particular, the calculation of ITCFs is further complicated by the need of multiplying
many exponentials of large matrices, see \eqref{the_boss}, which induces instabilities at large imaginary time.
This problem is already known in literature\cite{assaad,alhassid}.

\begin{figure*}
\centering
\includegraphics[scale=0.5]{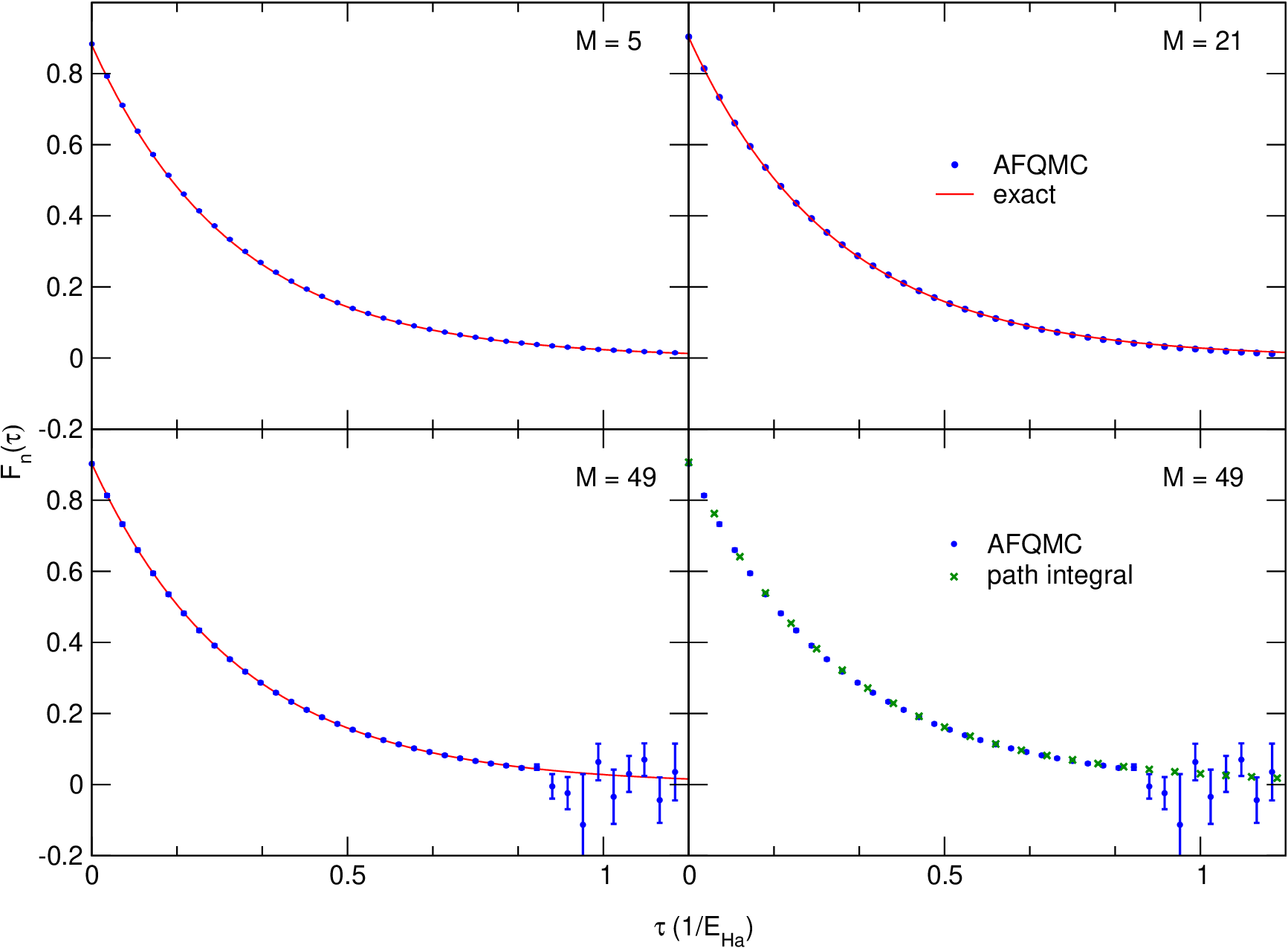}
\includegraphics[scale=0.5]{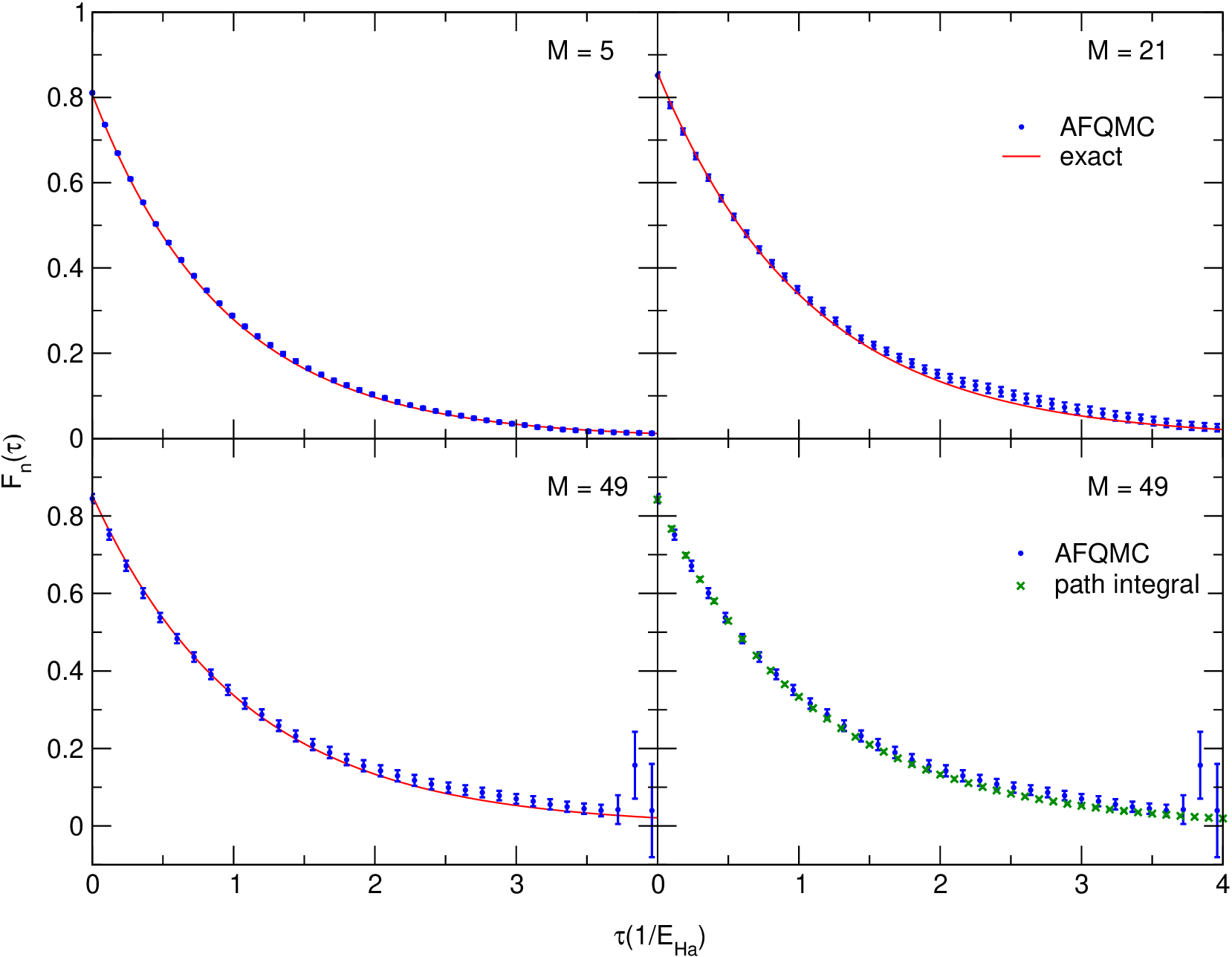}
\caption{(color online) Imaginary time correlation function of the momentum shift operator relative to $M = 5, 21,49$ at $r_s=1$ (upper panel) and $r_s=2$ (lower panel)
for $N_\uparrow = 1, N_\downarrow =1$. In the right lower box of each panel comparison between AFQMC and exact Path Integral QMC calculations is given.
When not visible, error bars are smaller than the symbol size.}
\label{ud_increase}
\end{figure*}

\begin{figure}
\centering
\includegraphics[scale=0.35]{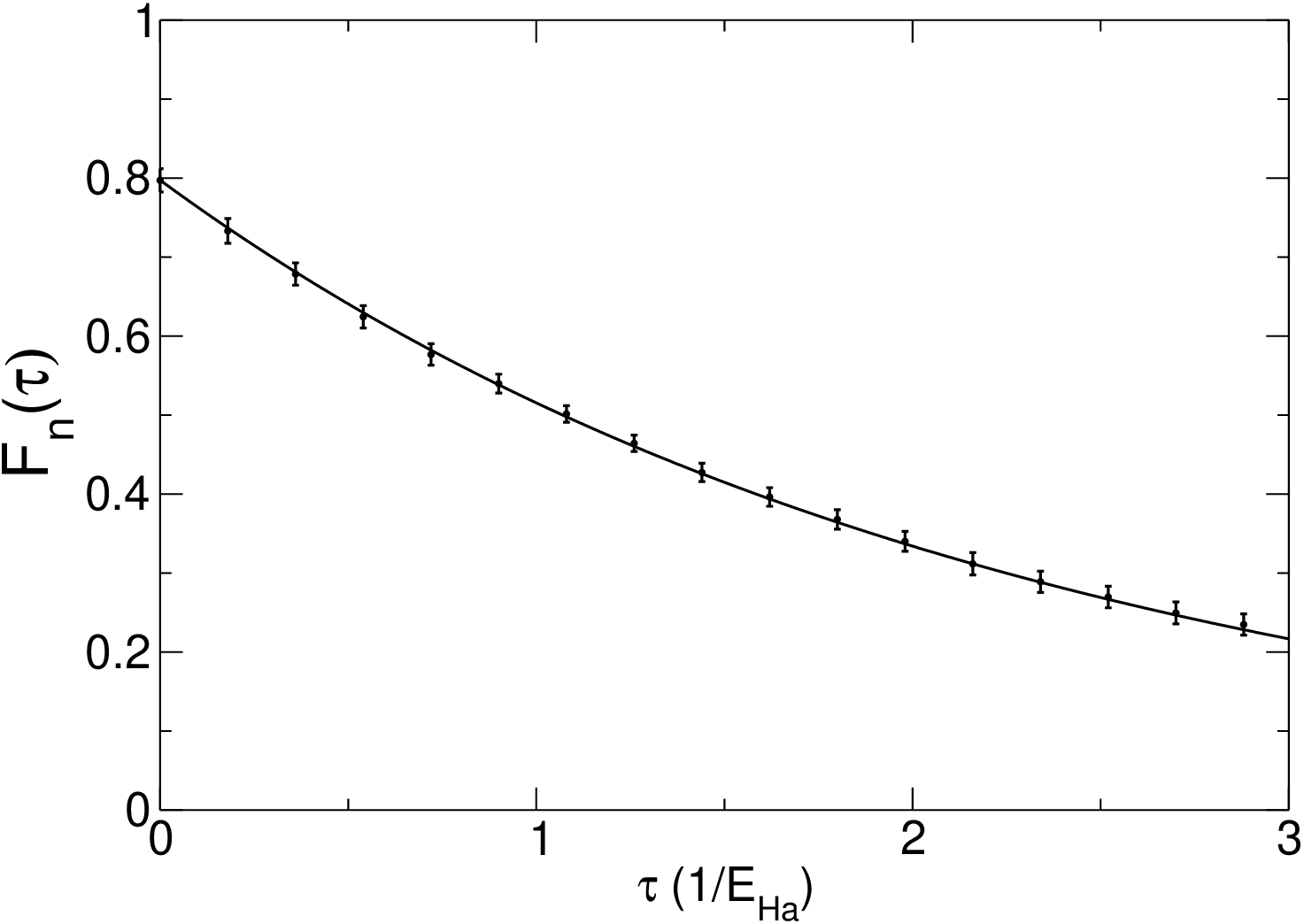}
\caption{Imaginary time correlation function of the momentum shift operator relative to $M = 21$ at $r_s=3$ for $N_\uparrow = 1, N_\downarrow =1$ (blue circles). 
Comparison with exact results (full line) is provided.} 
\label{ud_rs3}
\end{figure}

\begin{figure*}
\centering
\includegraphics[scale=0.5]{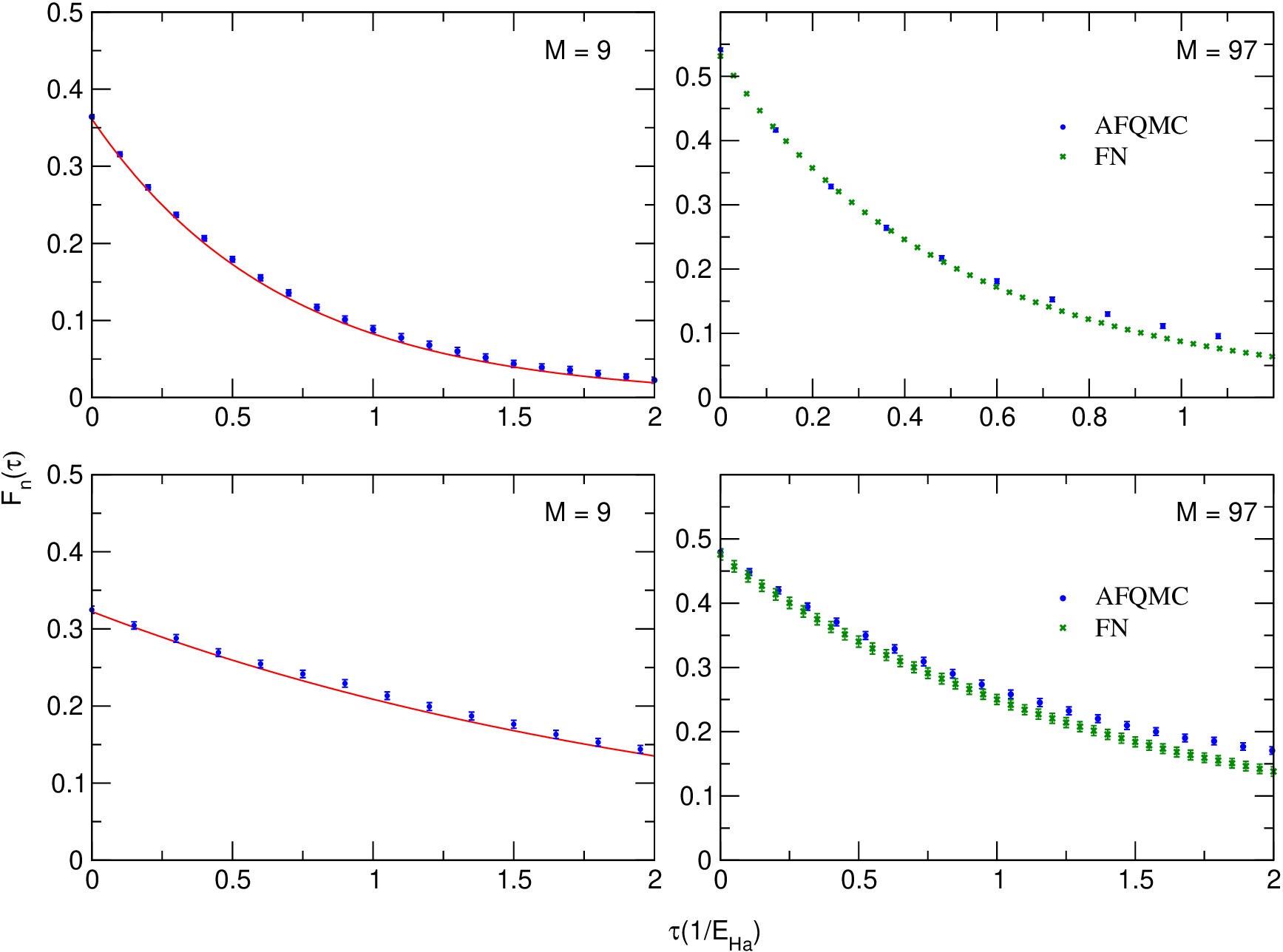}
\caption{(color online) Imaginary time correlation function of the momentum shift operator relative to $M = 9,97$ at $r_s=1$ (upper panel) and to $M = 13,97$ at $r_s=2$ (lower panel) for $N_\uparrow = 5, N_\downarrow =0$. In the right box of each panel comparison between AFQMC and FN calculations (green crosses) is given. When not visible, error bars are smaller than the symbol size.}
\label{5u_increase}
\end{figure*}

In Fig.~\ref{ud_increase} appear results relative to systems with $N_\uparrow = 1$, $N_\downarrow =1$, showing that AFQMC estimations 
of static and dynamic properties remain in satisfactory agreement with exact values even if $r_s$ and $M$ are raised respectively to $2$ and $49$.
For $M = 49$ we are also in good agreement with exact Path Integral QMC calculations, providing the exact result in the limit $M \to +\infty$, 
which cannot be explored via exact diagonalization.
The algorithm is able to reproduce exact values even at $r_s = 3$ and $M = 21$, as shown in Fig.~\ref{ud_rs3}.
We complete the study with calculations relative to systems with $N_\uparrow,N_\downarrow = (5,0)$. 
Results are shown in Fig.~\ref{5u_increase}.
For $M=9$ the quality of AFQMC calculations is still satisfactory, even if we observe a small overestimate of $F_{{\bf{n}}}(\tau)$, similar to that encountered in Fig.~\ref{ud_increase}.
Finally for $M = 97$ we compared our results with FN calculations. 
We observe good agreement between the estimates of the stastic property $F_{{\bf{n}}}(0)$ yield by both algorithms.
As far as finite $\tau$ ITCFs are concerned, We found that the discrepancy between the two results qualitatively resembles 
the discrepancy between exact solution and FN in the case of non interacting particles in Fig.~\ref{fnvsexact}: an encouraging result.

\pagebreak

\pagebreak

\pagebreak

\begin{table}
\centering
\begin{tabular}{c c c c c c c}
\hline\hline \\[0.5ex]
$N_\uparrow$ & $N_\downarrow$ & $r_s$ & $M$ & $\epsilon_0/N (AFQMC)$ & $\epsilon_0/N (exact)$ & $\braket{\Phi_0}{\Psi}$ \\ [1ex]
\hline \\ [0.5ex]
1 & 1 & 1.0 &  5 & -0.82255(5) & -0.82259 & 0.99999(5) \\ [1ex]
1 & 1 & 1.0 & 13 & -0.8315(1)  & -0.8313  & 0.9999(1)  \\ [1ex]
1 & 1 & 1.0 & 21 & -0.83338(6) & -0.83307 & 0.9989(7)  \\ [1ex]
1 & 1 & 1.0 & 49 & -0.83476(7) & -0.83441 & 0.9882(4)  \\ [1ex]
1 & 1 & 2.0 &  5 & -0.4282(1)  & -0.4282  & 0.9629(3)  \\ [1ex]
1 & 1 & 2.0 & 13 & -0.4351(1)  & -0.4330  & 0.9650(2)  \\ [1ex]
1 & 1 & 2.0 & 21 & -0.4359(3)  & -0.4339  & 0.9586(2)  \\ [1ex]
1 & 1 & 2.0 & 49 & -0.4362(3)  & -0.4345  & 0.9594(5)  \\ [1ex]
5 & 0 & 1.0 &  9 & 0.11327(2) & 0.11247 & 0.99185(1) \\ [1ex]
5 & 0 & 1.0 & 13 & 0.10726(3) & 0.10591 & 0.98600(7) \\ [1ex]
5 & 0 & 2.0 &  9 & -0.19485(1) & -0.19751 & 0.9863(4) \\ [1ex]
5 & 0 & 2.0 & 13 & -0.19878(2) & -0.20311 & 0.9683(3) \\ [1ex]
\hline
\end{tabular}
\label{ovr}
\caption{Exact (column 6) and calculated (column 5) ground state energy per particle in Hartree units, and overlap between exact and reconstructed ground state (column 7) for various systems (parameters are listed in columns 1 to 4).}
\end{table}


\pagebreak

\section{Conclusions}

In the present work we gave a detailed description of the phaseless AFQMC algorithm, we proposed a scheme for its application to the calculation of dynamical 
properties of zero temperature fermion systems and we tested the methodology against exact diagonalization for interacting few fermion systems.
Such tests revealed that it is actually possible to compute imaginary time correlation functions with satifactory accuracy, at least for systems with moderate number 
of particles and interaction strength. This is a very interesting result since it is known that there exist situations when the well
established and widely employed FN approximation scheme provides inaccurate results for ITCFs.
The present work indicates that AFQMC algorithm can become an important tool to calculate dynamical properties of few body systems
of experimental interest, like atomic or molecular systems.
Also the study of bulk systems is in principle feasible: a systematic work to reduce the complexity and
to improve numerical stability is however necessary. Numeric stabilization of the products of matrix exponentials involved in 
the calculation of imaginary time correlation functions could be realized with more refined linear algebra techniques 
\cite{gubernatis,alhassid}, enhancing the accuracy of the results.
These observations offer favorable prospects for the extended phaseless AFQMC method to be applied to larger electronic systems in future 
calculations, which will represent an occasion for detailed comparison with other QMC techniques.

\section{Acknowledgements}

This work has been supported by Regione Lombardia and CINECA Consortium through a LISA Initiative (Laboratory for Interdisciplinary Advanced Simulation) 2012 grant [http://www.hpc.cineca.it/services/lisa], and by a grant \emph{Dote ricerca}: FSE, Regione Lombardia.

\appendix

\section{Algorithmic Details}
\label{appA}

The aim of this appendix is completing the description of the extended AFQMC outlined in section I.

\subsection{Properties of Slater Determinants}
\label{appA1}

For a generic Slater determinant $\ket{\Psi}$ there exist single-particle orbitals $\ket{\psi_1} \dots \ket{\psi_N} \in \mathcal{H}$ for which $\ket{\Psi} = \ket{\psi_1 \dots \psi_N}_-$. As a consequence the state:
\begin{equation}
\label{slaterpar}
\begin{split}
\ket{\Psi} &= \sum_{i_1 \dots i_N} \braket{\varphi_{i_1}}{\psi_1} \dots \braket{\varphi_{i_N}}{\psi_N} \ket{\varphi_{i_1} \dots \varphi_{i_N}}_- = \\
&= \sum_{i_1 \dots i_N} \frac{\Psi_{i_1 1} \dots \Psi_{i_N N}}{\sqrt{N!}} \crt{i_1} \dots \crt{i_N} \qzero \\
\end{split}
\end{equation}
is completely and uniquely described by the $M \times N$ matrix $\Psi_{ij} = \braket{\phi_i}{\psi_j}$.

In the light of such parametrization it can be proved\cite{thouless} that for a generic $N$-particle Slater determinant $\ket{\Psi}$ and a generic one-body operator $\hat{O} = \sum_{i,j} \mathcal{O}_{ij} \crt{i} \dst{j}$ the state $e^{\hat{O}} \ket{\Psi}$ is still a Slater determinant, described by the matrix $e^{\mathcal{O}} \Psi$, so that the mainfold $\mathfrak{D}(N)$ of Slater determinants is closed under the action of exponentials of single-particle operators.

Equation \eqref{slaterpar} also enables the concrete calculation overlaps and matrix elements of one-body and two-body operators. In particular, if $\ket{\Psi}$, $\ket{\Phi}$ are generic non-orthogonal $N$-particle Slater determinants the following properties\cite{balian,senechal,thouless} hold:
\begin{equation}
\begin{split}
\braket{\Phi}{\Psi} &= \frac{\det[\Phi^\dagger \Psi]}{N!} \\
\frac{\braket{\Phi}{\crt{i} \dst{j}|\Psi}}{\braket{\Phi}{\Psi}} &= \left[ \Psi \left[ \Phi^\dag \Psi \right] \Phi^\dag \right]_{ji} = \mathcal{G}_{ij}\\
\frac{\braket{\Phi}{\crt{i} \crt{j}\dst{k} \dst{l}|\Psi}}{\braket{\Phi}{\Psi}} &= \mathcal{G}_{il} \mathcal{G}_{jk} - \mathcal{G}_{ik} \mathcal{G}_{jl} \\
\end{split}
\end{equation}

\subsection{The Hubbard-Stratonovich Transformation}
\label{appA2}

It is well known that the coefficients $\gamma_{ijlk}$ describing the interaction part of \eqref{hamiltonian} satisfy the relation $\gamma^*_{ijlk}=\gamma_{lkij}$ and can be consequently cast in a hermitian matrix $\Gamma_{(ki)(jl)} = \gamma_{ijlk}$ of order $M^2$. Due to the spectral theorem $\Gamma_{(ki)(jl)} = \sum_{\zeta=1}^{M^2} \mathcal{U}^*_{\zeta (ki)} \phantom{i} \Gamma_\zeta \phantom{i} \mathcal{U}_{\zeta (jl)}$ for some real-valued coefficients $\Gamma_\zeta$ and some unitary matrix $U$ of order $M^2$. As a consequence, \eqref{hamiltonian} can be put in the form:
\begin{equation}
\label{shiwei.dec}
\begin{split}
\hat{H} &= \hat{H}_0 - \frac{1}{2} \phantom{i} \sum_{\zeta=1}^{M^2} \phantom{i} \Gamma_\zeta \phantom{i} \left[ \frac{(\hat{O}_\zeta+\hat{O}^\dag_\zeta)^2}{2} + \frac{(i\hat{O}_\zeta-i\hat{O}^\dag_\zeta)^2}{2} \right] \\
\end{split}
\end{equation}
with:
\begin{equation}
\label{shiwei.dec.details}
\begin{split}
\hat{H}_0 &= \sum_{il} \phantom{i} \left[ \beta_{il} + \sum_j \gamma_{ijlj} \right] \, \hat{a}^\dag_{i} \hat{a}_{l} \\
\hat{O}_\zeta &= \sum_{jl} \phantom{i}U_{\zeta (jl)} \phantom{i} \hat{a}^\dag_{j} \hat{a}_{l} \\
\end{split}
\end{equation}
Notice that the interaction part of \eqref{shiwei.dec.details} has been replaced with a \emph{sum of squares of single-particle hermitian operators}. Inserting such expression in $\evodmc{\dt}$ and applying a Trotter-Suzuki decomposition:
\begin{equation}
\label{shiwei.dec.trotter}
\evodmc{\dt} = e^{\dt (\hat{H}_0-\epsilon_0)} \prod_{\zeta=1}^{M^2} e^{\frac{\dt}{2} \Gamma_\zeta (\hat{O}_\zeta+\hat{O}^\dag_\zeta)^2} e^{\frac{\dt}{2} \Gamma_\zeta (i\hat{O}_\zeta-i\hat{O}^\dag_\zeta)^2}
\end{equation}
To each of the factors appearing in \eqref{shiwei.dec.trotter} the Hubbard-Stratonovich Transformation applies, yielding \eqref{hs-transform} with:
\begin{equation}
\label{a.operator}
\hat{A}(\boldsymbol{\eta}) = \frac{\dt }{2} (\hat{H}_0-\epsilon_0) + \sum_\zeta \sqrt{\dt \Gamma_\zeta} \, (\eta_{1,\zeta}+i\eta_{2,\zeta}) \hat{O}_\zeta + h.c.
\end{equation}
which can be compactly written as:
\begin{equation}
\label{a.operator2}
\hat{A}(\boldsymbol{\eta}) = \frac{\dt }{2} (\hat{H}_0-\epsilon_0) - \sqrt{\dt} \, i \, \hat{\boldsymbol{B}} \cdot \boldsymbol{\eta}
\end{equation}

\subsection{The Importance Sampling Transformation}
\label{appA3}

We now explain in detail the derivation of equation \eqref{randomwalk}. First we introduce in the expression \eqref{stochastic2} $n$ arbitrary and possibly complex-valued shift parameters $\boldsymbol{\xi}_0 \dots \boldsymbol{\xi}_{n-1}$ obtaining:
\begin{eqnarray}
\label{hs-evolution}
\nonumber
\evodmc{n \dt} \psit \simeq \int &dg(\boldsymbol{\eta}_{n-1}-\boldsymbol{\xi}_{n-1}) \dots dg(\boldsymbol{\eta}_0-\boldsymbol{\xi}_0) \\ \nonumber
&\hat{G}(\boldsymbol{\eta}_{n-1}-\boldsymbol{\xi}_{n-1}) \dots \hat{G}(\boldsymbol{\eta}_0-\boldsymbol{\xi}_0) \psit \\ 
\end{eqnarray}
Then we recall that:
\begin{eqnarray}
dg(\boldsymbol{\eta}-\boldsymbol{\xi}) = dg(\boldsymbol{\eta}) \, e^{-\frac{\boldsymbol{\xi} \cdot \boldsymbol{\xi}}{2} - \boldsymbol{\eta} \cdot \boldsymbol{\xi}}
\end{eqnarray}
and obtain \eqref{randomwalk} inserting the identity:
\begin{equation}
\begin{split}
\label{product.of.ops}
&\, \, \gshift{n-1} \dots \gshift{0} \ket{\Psi_T} = \\
= &\frac{\gshift{n-1} \dots \gshift{0} \ket{\Psi_T}}{\braket{\Psi_T}{\gshift{n-1} \dots \gshift{0} \Psi_T}} \\
\prod_{k=0}^{n-1} &\frac{\braket{\Psi_T}{\gshift{k} \dots \gshift{0} \Psi_T}}{\braket{\Psi_T}{\gshift{k-1} \dots \gshift{0} \Psi_T}} \\
&\frac{\braket{\Psi_T}{\gshift{0} \Psi_T}}{\braket{\Psi_T}{\Psi_T}} = \\
= &\frac{\gshift{n-1} \dots \gshift{0} \ket{\Psi_T}}{\braket{\Psi_T}{\gshift{n-1} \dots \gshift{0} \Psi_T}} \\
&\mathfrak{W}\left[ \boldsymbol{\eta}_{n-1}, \boldsymbol{\xi}_{n-1} \dots \boldsymbol{\eta}_0, \boldsymbol{\xi}_0 \right] \\
\end{split}
\end{equation}
So far, the shift parameters are arbitrary. We subsequently fix their values to contain fluctuations in the importance function and therefore in the weight function. To this purpose, we expand $\hat{G}(\boldsymbol{\eta}-\boldsymbol{\xi})$ up to $\sqrt{\dt}$ obtaining:
\begin{equation}
\hat{G}(\boldsymbol{\eta}-\boldsymbol{\xi}) = \mathbb{I} - i \, \sqrt{\dt} \, (\boldsymbol{\eta}-\boldsymbol{\xi}) \cdot \hat{\boldsymbol{B}} + \mathcal{O}(\dt)
\end{equation}
Introducing this approximation in \eqref{importance} leads to:
\begin{equation}
\log\left[ \mathfrak{I} \left[ \boldsymbol{\eta},\boldsymbol{\xi};\ket{\Psi} \right] \right] = - \frac{|\boldsymbol{\xi}|^2}{2} + \boldsymbol{\eta} \cdot \boldsymbol{\xi} - i \, \sqrt{\dt} \, \frac{\braket{\Psi_T}{\boldsymbol{B}|\Psi}}{\braket{\Psi_T}{\Psi}} \cdot (\boldsymbol{\eta}-\boldsymbol{\xi})
\end{equation}
where the operation $\frac{\braket{\Psi_T}{\cdot|\Psi}}{\braket{\Psi_T}{\Psi}}$ shall be henceforth abbreviated with $\matel{\cdot}$.
Imposing $\partial_{\boldsymbol{\eta}} \log\left[ \mathfrak{I} \left[ \boldsymbol{\eta},\boldsymbol{\xi};\ket{\Psi} \right] \right] = 0$ fixes the value of the shift parameters to:
\begin{equation}
\label{opt.shift}
\boldsymbol{\xi}_{opt} = - i \sqrt{\dt} \matel{\boldsymbol{B}}
\end{equation}
Insertion of \eqref{opt.shift} into \eqref{importance} yields the stabilized expression for the importance function. A straightforward expansion of this quantity in powers of $\sqrt{\dt}$ leads to: 
\begin{equation}
\label{impo.opt}
\begin{split}
&\mathfrak{I} \left[ \boldsymbol{\eta}, \boldsymbol{\xi}_{opt} ; \ket{\Psi} \right] = 1 - \dt (\matel{H}-\epsilon_0) \\
&- \frac{\dt}{2} \left[ \mateldue{ \big| \boldsymbol{\eta} \cdot \left( \hat{\boldsymbol{B}} - \mateldue{\hat{\boldsymbol{B}}} \right) \big|^2 } - \mateldue{ \big| \hat{\boldsymbol{B}} - \matel{\boldsymbol{B}} \big|^2 } \right] + \mathcal{O}(\dt^{3/2}) \\
\end{split}
\end{equation}
The real local energy approximations \eqref{rle} is suggested by the observation that the term into square brackets in \eqref{impo.opt} has zero average over auxiliary field configurations,
and it consists in neglecting all terms of order $\dt$ in \eqref{impo.opt} except for the real part of $\matel{H}$. The imaginary part of $\matel{H}$ is neglected because it vanishes for $\ket{\Psi}$ equal tothe ground state.
 Empirical evidence shows that it is a reasonable approximation, but to our knowledge it is not supported by mathematical arguments.

\subsection{The Backpropagation Technique}
\label{appA4}

We now discuss the emergence of the backpropagated estimator \eqref{gs_average}. We express all imaginary time propagators appearing in \eqref{backpro} with \eqref{hs-evolution} and obtain the following representations for the numerator and the denominator:

\begin{equation}
\begin{split}
&\braket{\Psi_T}{\evodmc{(m+n)\dt}|\Psi_T} = \\
&\int dg(\boldsymbol{\eta}_{m+n-1}) \dots dg(\boldsymbol{\eta}_0) \braket{\Psi_T}{ \prod_{i=0}^{m+n-1} \gofeta{i} |\Psi_T} \\
&\braket{\Psi_T}{\evodmc{m \dt} \hat{O} \evodmc{n \dt}|\Psi_T} = \\
&\int dg(\boldsymbol{\eta}_{m+n-1}) \dots dg(\boldsymbol{\eta}_0) \braket{\Psi_T}{ \prod_{i=n}^{m+n-1} \gofeta{i} \hat{O} \prod_{i=0}^{n-1} \gofeta{i} |\Psi_T} \\
\end{split}
\end{equation}
where the symbol $\prod_{i=i_1}^{i_2} \gofeta{i}$ stands for the product $\gofeta{i_2} \dots \gofeta{i_1}$. Further application of the importance sampling transformation and of identity \eqref{product.of.ops} yields:
\begin{equation}
\begin{split}
&\braket{\Psi_T}{\evodmc{(m+n)\dt}|\Psi_T} = \\
&\int dg(\boldsymbol{\eta}_{m+n-1}) \dots dg(\boldsymbol{\eta}_0) \mathfrak{W}\left[ \boldsymbol{\eta}_{m+n-1}, \boldsymbol{\xi}_{m+n-1} \dots \boldsymbol{\eta}_0, \boldsymbol{\xi}_0 \right] \\ 
&\braket{\Psi_T}{\evodmc{m \dt} \hat{O} \evodmc{n \dt}|\Psi_T} = \\
&\int dg(\boldsymbol{\eta}_{m+n-1}) \dots dg(\boldsymbol{\eta}_0) \mathfrak{W}\left[ \boldsymbol{\eta}_{m+n-1}, \boldsymbol{\xi}_{m+n-1} \dots \boldsymbol{\eta}_0, \boldsymbol{\xi}_0 \right] \\ 
&\frac{\braket{\Psi_T}{ \prod_{i=n}^{m+n-1} \gshift{i} \hat{O} \prod_{i=0}^{n-1} \gshift{i} |\Psi_T}}{\braket{\Psi_T}{ \prod_{i=0}^{m+n-1} \gshift{i}|\Psi_T}} \\
\end{split}
\end{equation}
the estimator for which is obviously \eqref{gs_average}. Notice that {\emph{the same weights appearing in}} \eqref{flow} are involved in \eqref{gs_average}. 

\subsection{The phaseless AFQMC estimator for ITCFs}
\label{appA5}

We now explain in detail the derivation of equations \eqref{ext_phaseless} and \eqref{the_boss}. The last passage of \eqref{ext_phaseless} is a manipulation of the operator product $\gofeta{n-1} \dots \gofeta{n-r} \crt{k} \dst{l} $. First, we observe that if $\hat{A} = \sum_{i,j} \mathcal{A}_{i,j} \crt{i} \dst{j}$ is a one-body operator:
\begin{equation}
\begin{split}
e^{\tau\hat{A}} \crt{k} e^{-\tau\hat{A}} &= \sum_i \left[ e^{\tau\mathcal{A}} \right]_{ik} \crt{i} \\ 
e^{\tau\hat{A}} \dst{l} e^{-\tau\hat{A}} &= \sum_j \left[ e^{-\tau\mathcal{A}} \right]_{lj} \dst{j} \\
\end{split}
\end{equation}
As an immediate consequence:
\begin{equation}
\label{propagator.swap}
e^{\tau\hat{A}} \crt{k} \dst{l} = \sum_{ij} \left[ e^{\tau\mathcal{A}}\right]_{ik} \crt{i} \dst{j} \left[e^{-\tau\mathcal{A}}\right]_{lj} \, e^{\tau\hat{A}}
\end{equation}
showing that the exponential of a one-body operator can be moved to the right of a product $\crt{k} \dst{l}$ at the cost of introducing the matrix $e^{\tau\mathcal{A}}$ and its inverse.
Iterated application of formula \eqref{propagator.swap} to the operator product $\gofeta{n-1} \dots \gofeta{n-r} \crt{k} \dst{l} $ yields:
\begin{equation}
\begin{split}
&\gofeta{n-1} \dots \gofeta{n-r} \crt{k} \dst{l} = \\
= &\sum_{ij} \left[ e^{\mathcal{A}(\boldsymbol{\eta}_{n-r})} \dots e^{\mathcal{A}(\boldsymbol{\eta}_{n-1})} \right]_{ik} \crt{i} \dst{j} \left[ e^{-\mathcal{A}(\boldsymbol{\eta}_{n-1})} \dots e^{-\mathcal{A}(\boldsymbol{\eta}_{n-r})} \right]_{lj} \cdot \\
\cdot &\gofeta{n-1} \dots \gofeta{n-r} \\
\end{split}
\end{equation}
and justifies the last passage of equation \eqref{ext_phaseless}. To obtain \eqref{the_boss} we observe, as in the backpropagation technique, that:
\begin{equation}
\label{fqt.estimator}
\begin{split}
&F_{\hat{A},\hat{B}}(r \dt) \simeq \frac{1}{N} \frac{1}{\bra{\Psi_T}\evodmc{(m+n-r)\dt}\psit} \cdot \\
\cdot &\braket{\Psi_T}{\evodmc{m \dt} \hat{A} \evodmc{r\dt}\hat{B} \evodmc{(n-r) \dt}|\Psi_T} \\
\end{split}
\end{equation}
notice that, unlike in \eqref{backpro}, at the denominator of the previous equation only $m+n-r$ integrations over auxiliary fields configurations are involved. Expressing all imaginary time propagators appearing in \eqref{fqt.estimator} with \eqref{hs-evolution}, recalling \eqref{ext_phaseless} and applying the importance sampling transformation to both numerator and denominator of the previous equation lead to:
\begin{equation}
\label{ITCF.den}
\begin{split}
&\braket{\Psi_T}{\evodmc{(m+n-r)\dt}|\Psi_T} = \\
&\int dg(\boldsymbol{\eta}_{m+n-1}) \dots dg(\boldsymbol{\eta}_n) dg(\boldsymbol{\eta}_{n-r}) \dots dg(\boldsymbol{\eta}_0) \\
&\braket{\Psi_T}{ \prod_{i=n}^{m+n-1} \gofeta{i} \, \prod_{i=0}^{n-r-1} \gofeta{i} |\Psi_T} \\
\end{split}
\end{equation}
\begin{equation}
\label{ITCF.num}
\begin{split}
&\braket{\Psi_T}{\evodmc{m \dt} \hat{A} \evodmc{r\dt} \hat{B} \evodmc{n \dt}|\Psi_T} = \\
&\sum_{ijkl} \mathcal{B}_{kl} \int dg(\boldsymbol{\eta}_{m+n-1}) \dots dg(\boldsymbol{\eta}_0) \\
&\braket{\Psi_T}{ \prod_{i=n}^{m+n-1} \gofeta{i} \hat{A} \crt{i}\dst{j} \prod_{i=0}^{n-1} \gofeta{i} |\Psi_T} \\
&\mathcal{D}(\boldsymbol{\eta}_{n-1}\dots \boldsymbol{\eta}_{n-r})_{ik} \mathcal{D}^{-1}(\boldsymbol{\eta}_{n-1}\dots \boldsymbol{\eta}_{n-r})_{lj}\\
\end{split}
\end{equation}
Further application of the importance sampling transformation and of identity \eqref{product.of.ops} yields:
\begin{equation}
\begin{split}
\label{ITCF.den.2}
&\braket{\Psi_T}{\evodmc{(m+n-r)\dt}|\Psi_T} = \\
&\int dg(\boldsymbol{\eta}_{m+n-1}) \dots dg(\boldsymbol{\eta}_n) dg(\boldsymbol{\eta}_{n-r}) \dots dg(\boldsymbol{\eta}_0) \\
&\mathfrak{W}\left[ \boldsymbol{\eta}_{m+n-1}, \boldsymbol{\xi}_{m+n-1} \dots \boldsymbol{\eta}_n, \boldsymbol{\xi}_n \boldsymbol{\eta}_{n-r}, \boldsymbol{\xi}_{n-r} \dots \boldsymbol{\eta}_0, \boldsymbol{\xi}_0 \right] \\
\end{split}
\end{equation}
\begin{equation}
\label{ITCF.num.2}
\begin{split}
&\braket{\Psi_T}{\evodmc{m \dt} \hat{A} \evodmc{r\dt} \hat{B} \evodmc{n \dt}|\Psi_T} = \\
&\sum_{ijkl} \mathcal{B}_{kl} \int dg(\boldsymbol{\eta}_{m+n-1}) \dots dg(\boldsymbol{\eta}_0) \\
&\mathfrak{W}\left[ \boldsymbol{\eta}_{m+n-1}, \boldsymbol{\xi}_{m+n-1} \dots \boldsymbol{\eta}_0, \boldsymbol{\xi}_0 \right] \\
&\frac{\braket{\Psi_T}{ \prod_{i=n}^{m+n-1} \gshift{i} \hat{A} \crt{i}\dst{j} \prod_{i=0}^{n-1} \gshift{i} |\Psi_T}}{ \braket{\Psi_T}{ \prod_{i=0}^{m+n-1} \gshift{i} |\Psi_T} } \\
&\mathcal{D}(\boldsymbol{\eta}_{n-1} - \boldsymbol{\xi}_{n-1} \dots \boldsymbol{\eta}_{n-r} - \boldsymbol{\xi}_{n-r})_{ik} \\
&\mathcal{D}^{-1}(\boldsymbol{\eta}_{n-1} - \boldsymbol{\xi}_{n-1} \dots \boldsymbol{\eta}_{n-r} - \boldsymbol{\xi}_{n-r} )_{lj}\\
\end{split}
\end{equation}
an estimator for which is precisely \eqref{the_boss}. Notice that the weights appearing in the denominator of \eqref{the_boss} are the same appearing in \eqref{gs_average}, whereas at the denominator {\em other weights appear}, which are constructed with a slightly modified recursion relation:
\begin{equation}
\mathfrak{W}^{(w)}_{k+1} = 
\begin{cases}
\mathfrak{W}^{(w)}_{k} \quad \quad \quad \quad \quad \quad \mbox{if $n-r \leq k \leq n-1$}\\
\\
\mathfrak{W}^{(w)}_{k} \, \mathfrak{I}\left[ \boldsymbol{\eta}^{(w)}_k , \boldsymbol{\xi}^{(w)}_k ; \ket{\Psi^{(w)}_{k}} \right] \quad \mbox{otherwise}\\
\end{cases}
\end{equation}

$ $

\section{ITCFs for the Ideal Fermi Gas}
\label{appC}

In the case of a non-interating system the ITCF $F_{{\bf{q}}}(\tau)$ takes the form:
\begin{equation}
\label{it.ideal.1}
\begin{split}
F_{{\bf{q}}}(\tau) &= \frac{1}{N} \braket{\Psi_0|\hat{\rho}_{\boldsymbol{q}}(\tau) \hat{\rho}_{-\boldsymbol{q}}}{\Psi_0} = \\
&= \frac{1}{N} \sum_{\boldsymbol{p},\boldsymbol{p}'} \sum_{\sigma\sigma'} \braket{\Psi_0|\hat{a}^\dag_{\boldsymbol{p}-\boldsymbol{q},\sigma}(\tau) \hat{a}_{\boldsymbol{p},\sigma}(\tau) \hat{a}^\dag_{\boldsymbol{p}'+\boldsymbol{q},\sigma'} \hat{a}_{\boldsymbol{p}',\sigma'}}{\Psi_0}
\end{split}
\end{equation}
For a spin polarized system, using Heisenberg representation and Wick's theorem, formula \eqref{it.ideal.1} can be reduced to:
\begin{equation}
\label{it.ideal.2}
F_{{\bf{q}}}(\tau) = \frac{1}{N} \sum_{\boldsymbol{p}} e^{- \tau (\epsilon_{\boldsymbol{p}}-\epsilon_{\boldsymbol{p}-\boldsymbol{q}})} \, \Theta(k_F - |\boldsymbol{p}-\boldsymbol{q}|) \, \Theta(|\boldsymbol{p}|-k_F)
\end{equation}
Numeric evaluation of \eqref{it.ideal.2} yields $F_{{\bf{q}}}(\tau)$. For $N_\uparrow=5$, $N_\downarrow=0$, $r_s = 1$ and $\boldsymbol{q} = \frac{2\pi}{L} (1,0)$ the nonvanishing contributions to \eqref{it.ideal.2} come from $\boldsymbol{p} = \frac{2\pi}{L} (1,0), \frac{2\pi}{L}(0,1), \frac{2\pi}{L}(0,-1)$. Consequently:
\begin{equation}
\label{it.ideal.3}
F_{{\bf{q}}}(\tau) = \frac{e^{-\frac{6 \pi}{5} \tau}}{5} + \frac{2 e^{-\frac{2 \pi}{5} \tau}}{5}
\end{equation}

\end{document}